\documentclass{article}

\usepackage{arxiv}

\usepackage[section]{placeins}

\usepackage[utf8]{inputenc} 
\usepackage[T1]{fontenc}    
\usepackage{hyperref}       
\usepackage{url}            
\usepackage{booktabs}       
\usepackage{amsfonts}       
\usepackage{nicefrac}       
\usepackage{microtype}      
\usepackage{lipsum}
\usepackage{graphicx}
\usepackage{amsmath}
\usepackage{float}
\usepackage{booktabs,tabularx,array}
\usepackage{subcaption}
\usepackage[numbers]{natbib}

\graphicspath{ {./images/} }

\title{Intelligent Shanghai Typhoon Model (ISTM): A generative probabilistic emulator for typhoon hybrid modeling}

\begin{document}

\maketitle

\begin{center}
    Zeyi Niu$^{1,2}$,
    Wei Huang$^{1,}$\footnote{Corresponding Author: Wei Huang at huangw@typhoon.org.cn},
    Sirong Huang$^{3}$,
    Bo Qin$^{2}$,
    Mengqi Yang$^{1}$,
    Haofei Sun$^{1}$,
    Zhaoyang Huo$^{1}$,
    Haixia Xiao$^{2}$\\
    \vspace{0.5cm}
    
    $^{1}~$\textit{Shanghai Typhoon Institute, and Key Laboratory of Numerical Modeling for Tropical Cyclone of the China Meteorological Administration, Shanghai, China}\\
    $^{2}~$\textit{Department of Atmospheric and Oceanic Sciences and Institute of Atmospheric Sciences, Fudan University, Shanghai 200438, China}\\
    $^{3}~$\textit{Shanghai Engineering Research Center of Intelligent Education and Bigdata, Shanghai Normal University, Shanghai, China}


\end{center}

\vskip 0.25cm

\begin{abstract}
To address the systematic underestimation of typhoon intensity in artificial intelligence weather prediction (AIWP) models, we propose the Intelligent Shanghai Typhoon Model (ISTM)—a unified regional-to-typhoon generative probabilistic forecasting system based on a two-stage UNet–Diffusion framework. ISTM learns a downscaling mapping from 4 years of 25 km ERA5 reanalysis to a 9 km high resolution typhoon reanalysis dataset, enabling the generation of kilometer-scale near-surface variables and maximum radar reflectivity from coarse-resolution fields. The evaluation results show that the two-stage UNet–Diffusion model significantly outperforms both ERA5 and the baseline UNet regression in capturing the structure and intensity of surface winds and precipitation. After fine-tuning, ISTM can effectively map AIFS forecasts, an advanced AIWP model, to high-resolution forecasts from AI-physics hybrid Shanghai Typhoon Model, substantially enhancing typhoon intensity predictions while preserving track accuracy. This positions ISTM as an efficient AI emulator of hybrid modeling system, achieving fast and physically consistent downscaling. The proposed framework establishes a unified pathway for the co-evolution of AIWP and physics-based numerical models, advancing next-generation typhoon forecasting capabilities.
\end{abstract}


\section{Introduction} 
In recent years, artificial intelligence weather prediction (AIWP) models have rapidly evolved from data-driven approaches (Bi et al., 2023; Lam et al., 2023; Chen et al., 2023; Lang et al., 2023), to ensemble-based AIWP models (Price et al., 2025), AI-enabled data assimilation systems (Xu et al., 2024), end-to-end forecasting systems (Allen et al., 2025), and even foundation models exemplified by Aurora, which are designed to generate globally consistent, multivariable atmospheric forecasts using end-to-end AI architectures (Bodnar et al., 2025). From these developments, the meteorological research community has gradually converged on several key insights: (i) AIWP models have surpassed state-of-the-art numerical weather prediction models (NWPs) like ECMWF HRES in predicting large-scale atmospheric circulation, particularly in Z500 anomaly correlation and typhoon track forecasting; (ii) however, these models still struggle to forecast extreme events, especially tropical cyclone intensity, which is systematically underestimated. This underestimation has been attributed to both insufficient training samples (Sun et al., 2025) and the use of mean squared error (MSE) loss functions, which bias predictions toward smoother outcomes (Subich et al., 2025). To mitigate this intensity underestimation, several recent studies have used AIWP model forecasts to constrain (i.e., nudge) high-resolution numerical weather prediction models to produce more skillful typhoon forecasts. For example, Liu et al. (2024) integrated the Pangu-Weather model with a high-resolution ocean–atmosphere coupled model to improve tropical cyclone intensity prediction. Similarly, Environment and Climate Change Canada (ECCC) achieved substantial improvements by combining their global GEM model with the GraphCast model (Husain et al., 2024). Niu et al. (2025) developed a real-time hybrid typhoon forecasting system by coupling AIWP models with the Shanghai Typhoon Model (SHTM). These AI–physics hybrid frameworks are emerging as a new operational paradigm in global NWP centers. However, they still depend on physics model integrations, which remain computationally expensive, and the spectral nudging process typically increases runtime by 20\%–30\%.

This study addresses a critical challenge: how to substantially improve the computational efficiency of AI–physics hybrid models without compromising their forecast accuracy. Since the hybrid framework essentially performs a physically based dynamical downscaling of AIWP forecasts, we propose learning this process via neural networks to emulate it directly. Existing AI-based downscaling studies have already shown promise (Lian et al., 2024). Generative approaches, particularly Generative Adversarial Networks (GANs) (Leinonen et al., 2020; Price and Rasp, 2022) and diffusion models (Hatanaka et al., 2023; Lopez-Gomez et al., 2025), have emerged as state-of-the-art techniques. For example, NVIDIA’s CorrDiff model (Mardani et al., 2025) combines a UNet for mean regression with a diffusion model to generate high-resolution forecasts from coarse ERA5 inputs. However, CorrDiff is limited to a small region centered over Taiwan and cannot simulate long-duration typhoon evolution. Other efforts have built regional AIWP models from reanalysis data, such as the layered graph-based architecture by Oskarsson et al. (2025) and boundary-aware models like Yinglong (Xu et al., 2025). Yet, these models are not trained for the East Asian domain and offer limited lead times, rendering them insufficient for typhoon applications. Regardless of model type, the quality of initial conditions and training datasets remains essential for forecast skill. Therefore, this study constructs a high-resolution, typhoon-focused reanalysis dataset for the Asia–Pacific domain, based on the operational hybrid SHTM, to serve as training data for an AI emulator. Training this dataset, we develop the Intelligent Shanghai Typhoon Model (ISTM), a generative, two-stage UNet–Diffusion emulator that learns the downscaling relationship between coarse-resolution ERA5 and high-resolution typhoon reanalysis. The ISTM enables efficient and accurate emulation of SHTM forecasts, representing a critical step toward accelerating model forecasts and advancing a unified framework for next-generation typhoon prediction.

\section{Method and Data}
\subsection{SHTM configurations and developments}
SHTM is the operational mesoscale regional forecasting system of the Shanghai Meteorological Service (SMS), launched for real-time use in 2023 as a core component of the Shanghai Warning and Risk Model System (SWARMS). It is specifically designed for typhoon prediction. SHTM is built upon the Weather Research and Forecasting (WRF) model (version 4.3), coupled with the Gridpoint Statistical Interpolation (GSI) data assimilation system (version 3.7). The model is initialized twice daily at 0000 UTC and 1200 UTC, and produces forecasts up to 120 hours. SHTM operates on a 953×701 horizontal grid with 9-km resolution and 56 eta vertical levels extending to 10 hPa. The model domain spans 0.5°S–61.5°N latitude and 58.3°E–172.7°E longitude (purple curves in Fig. 2b), encompassing mainland China, surrounding land areas, and the western North Pacific, a key region for typhoon activity. The model’s physics configuration includes the Thompson microphysics scheme, the multi-scale Kain–Fritsch cumulus parameterization, the RRTMG radiation scheme (for both longwave and shortwave), the Noah land surface model, and the Yonsei University planetary boundary layer scheme. 

To overcome the limitations of standalone AIWP and physics-based models, we adopt a hybrid modeling strategy that integrates the large-scale forecasts of AIWP models with the mesoscale capabilities of the SHTM using spectral nudging. This hybrid system combines the strengths of AI-based synoptic forecasting and physically consistent regional modeling. In 2024, this approach was operationalized by coupling FuXi with SHTM, forming the Hybrid SHTM–FuXi system (Niu et al., 2025b) \citep{Niu2025b}. In 2025, the framework is upgraded by replacing FuXi with the newly released AIFS model, which provides higher-frequency forecasts and improved large-scale performance. The resulting SHTM system serves as a high-quality reference for typhoon track and intensity prediction and forms the basis for developing our AI-based downscaling emulator that replicates hybrid forecasts at a fraction of the computational cost.
\subsection{SHTM and AIWP evaluations}
In recent years, the China Meteorological Administration (CMA) has initiated trial operations of AIWP models at the National Meteorological Centre (NMC). To assess their value in operational typhoon forecasting, the Shanghai Typhoon Institute (STI) evaluated six AIWP models—AIFS, GraphCast, Fengwu, Fuxi, and Pangu—using the official CMA verification protocol. Forecast errors for tropical cyclone track and intensity during the 2024 season were obtained from the CMA Tropical Cyclone Data Center and compared with those from the original SHTM and its hybrid version in 2024. All AIWP models are initialized from 0.1° ECMWF HRES fields. Figure 1 shows that hybrid SHTM consistently outperforms all models in both track and intensity forecasts, especially within 96 h lead times. In contrast, most AIWP models significantly underestimate typhoon intensity. While ECMWF-IFS also shows systematic underestimation, hybrid SHTM achieves the best overall performance. To enable efficient use of such high-skill forecasts, this study develops an AI-based emulator that downscales AIWP fields into high-resolution outputs resembling Hybrid SHTM forecasts. 

\begin{figure}[htbp]
    \centering
    \includegraphics[width=0.8\textwidth]{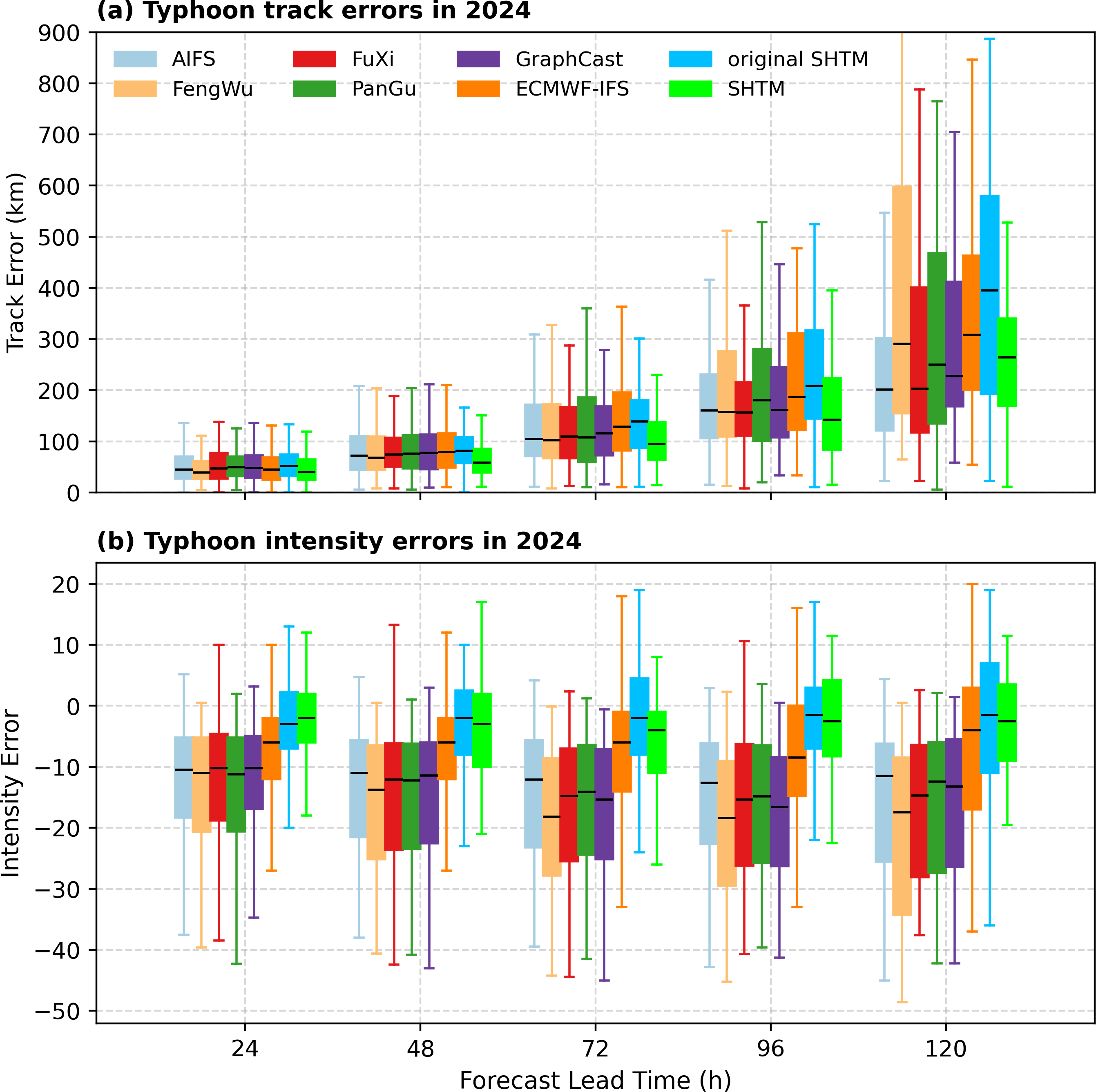}
    \caption{Boxplots of (a) track errors (unit: km) and (b) intensity errors (unit: m/s) from different models, including mainstream ML weather prediction models, ECMWF-IFS, original SHTM and SHTM, evaluated by all WNP typhoons in 2024.}
    \label{fig:fig1}
\end{figure}

\subsection{High-Resolution regional typhoon reanalysis}
This study develops a four-year high-resolution regional reanalysis dataset (HiRes) focused on typhoon intensity and structural evolution, based on the Shanghai Typhoon Model. For each year, a continuous year-long forecast is initialized at the beginning of the calendar year, with outputs saved every 6 hours at a horizontal resolution of 9~km. During model integration, spectral nudging is applied every 6 hours to align the large-scale background flow with ERA5 reanalysis, and tropical cyclone structure is improved via assimilation of Tropical Cyclone Vital Statistics Records (TCVitals), which include minimum pressure, maximum wind speed, storm size, and storm shape.To ensure consistency with the synoptic-scale environment, spectral nudging is applied only to the large-scale components of prognostic variables. As shown in Fig.~2a, for a given variable~$F$, the nudged field is defined as:
\[
F^{\mathrm{nud}} = F^{\mathrm{SHTM}} + \omega \left( F^{\mathrm{ERA5}} - F^{\mathrm{SHTM}} \right)_{\mathrm{LS}},
\]
where $F^{\mathrm{SHTM}}$ is the native model prediction, $F^{\mathrm{ERA5}}$ is the ERA5 analysis, $\omega \in [0, 1]$ is the relaxation coefficient, and the subscript ``LS'' denotes the large-scale spectral components.This formulation allows HiRes to preserve small-scale structures generated by the SHTM physics while maintaining consistency with the large-scale flow of ERA5. Specifically, spectral nudging is applied to horizontal wind components ($U$, $V$) and temperature ($T$), while relative humidity is excluded based on previous findings that it may degrade typhoon intensity \citep{Husain2024}. The spectral truncation targets planetary-scale wavelengths ($>1000$~km), using zonal and meridional wavenumbers of 8 and 7, respectively. The relaxation timescale~$\tau$ is set to six hours, matching the update frequency of ERA5 and avoiding artifacts from temporal interpolation. More details can be found in \citet{Niu2025}.

The effectiveness of the HiRes dataset is exemplified by its performance during Typhoon Doksuri (2023). Figure 2 compares the 10-m wind fields from ERA5 and HiRes. While ERA5 exhibits a diffuse and weakened wind structure (Fig. 2c), HiRes captures a more compact and intense typhoon vortex (Fig. 2d). Furthermore, the HiRes wind fields show excellent agreement with spaceborne Synthetic Aperture Radar (SAR) observations (Fig. 2e), particularly in the representation of outer wind band intensity and structure. These improvements in structural realism are critical for downstream applications such as model verification and training of AI-based forecasting systems.

\begin{figure}[htbp]
    \centering
    \includegraphics[width=0.8\textwidth]{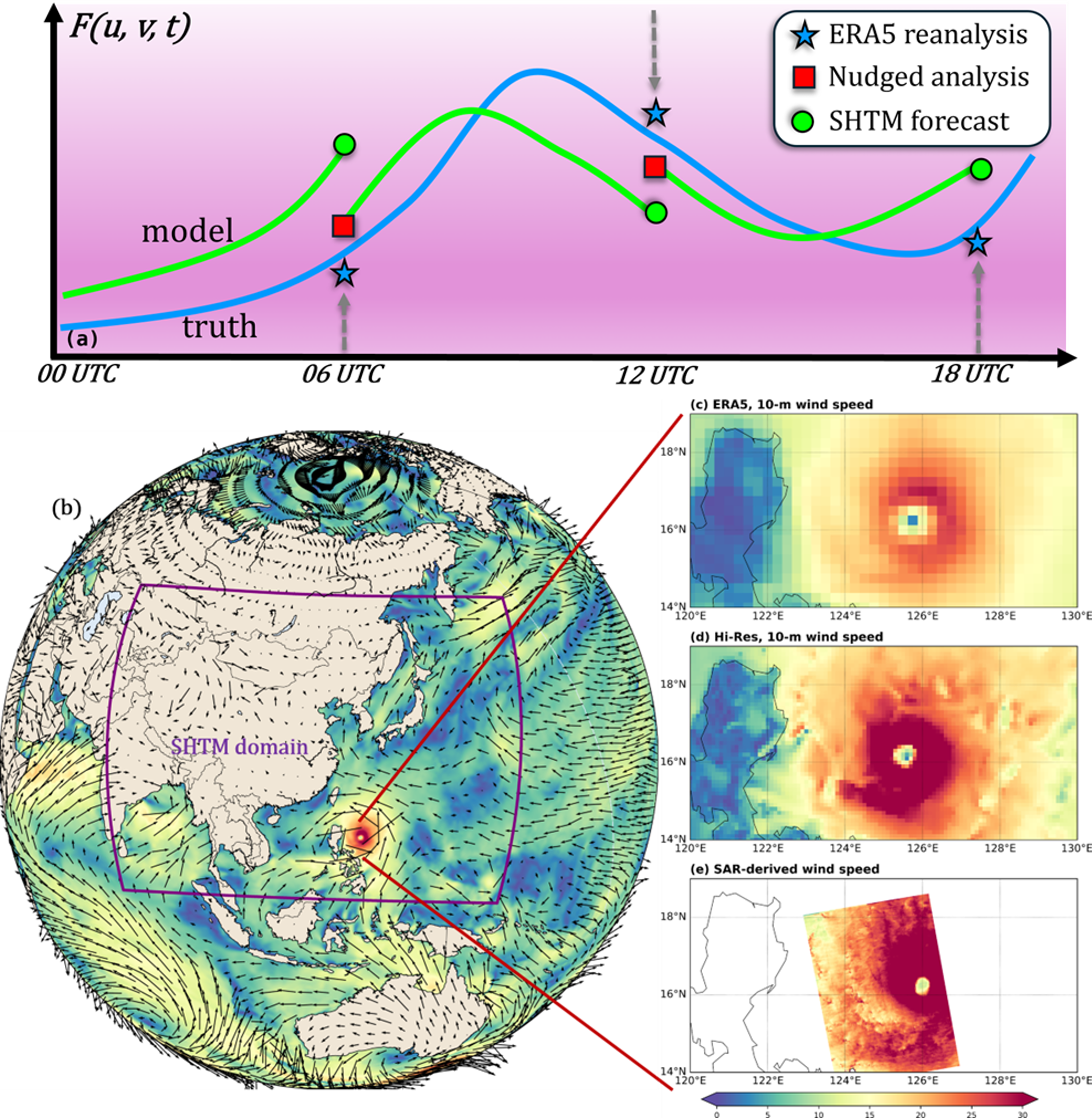}
    \caption{(a) Conceptual illustration of the spectral nudging process, depicting the evolution of forecast variables over UTC times. The blue lines represent the true states, the green circles denote an unconstrained SHTM forecasts, red squares indicate the analyzed states after nudging, and blue stars represent large scale information from ERA5. (b) Spatial distribution of the global 10-m wind of ERA5, (c)-(e) ERA5, HiRes, and SAR wind speed from the selected area, around 1200 UTC 24 July 2023 during Typhoon Doksuri (2023). The purple curves stand for the SHTM domain.}
    \label{fig:fig2}
\end{figure}

\begin{figure}[htbp]
    \centering
    \includegraphics[width=0.8\textwidth]{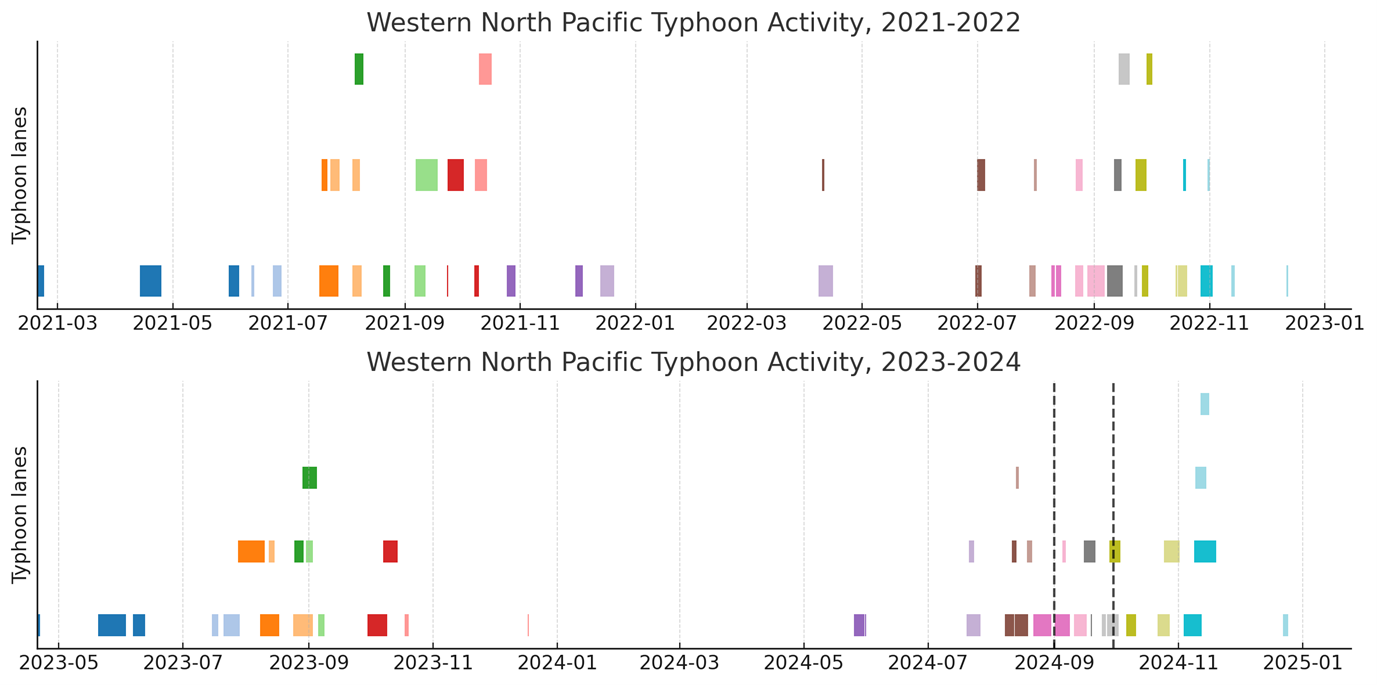}
    \caption{Western North Pacific typhoon activity from 2021 to 2024. Horizontal bars show each named typhoon's lifetime from genesis to dissipation; bar length is proportional to duration. Bars are packed into nonoverlapping “typhoon lanes” to indicate concurrent events. The data for September 2024 (between the black dashed lines) is used as the test dataset.}
    \label{fig:fig3}
\end{figure}

\section{UNet–Diffusion architecture and training Setup}
We adopt a two-step downscaling approach: a UNet-based deterministic regression first predicts the conditional mean of HiRes targets, followed by a conditional diffusion model that learns and generates the residual field, enabling reconstruction of physically realistic high-resolution structures.
\subsection{UNet regression}
In the first stage, we perform deterministic regression of the target HiRes field $Y$ to parametrically estimate its conditional mean, denoted by $\mu_{\theta}$. We employ a UNet-style architecture comprising six encoder--decoder stages, with ResNet-based blocks serving as the backbone. Each encoder stage progressively reduces spatial resolution to capture high-level semantic features, while the decoder restores the original resolution, facilitating precise fine-grained detail recovery. Skip connections link corresponding encoder and decoder stages, enabling the fusion of fine-grained details and high-level semantic information. The model is optimized using mean squared error (MSE) minimization, ensuring convergence toward the conditional mean (Wang et al., 2018; Lee et al., 2024).
\subsection{Residual denoising diffusion}
The second stage refines the parametric conditional mean $\mu_{\theta}$ probabilistically using a conditional diffusion model (CDM) that learns the conditional distribution of the residual
$R_{\theta} = Y - \mu_{\theta}$. As illustrated in Fig.~4c, the CDM applies a forward diffusion process over $T$ steps to progressively perturb the residual into standard Gaussian noise. A denoising network, conditioned on the upsampled ERA5 fields and $\mu_{\theta}$, is trained to reverse this process.

In this study, the denoising network adopts a four-stage encoder--decoder UNet architecture. As shown in Fig.~4d, each backbone block contains two ResNet units with spatial attention and explicit down-/up-sampling layers. The diffusion time step $t$ is embedded via a sinusoidal positional encoder and then projected through a multilayer perceptron (MLP), producing two modulation vectors $\gamma_t$ and $\beta_t$ that control feature-wise scaling and shifting within the denoising network. At inference, the CDM draws a sample from standard Gaussian noise and iteratively applies the trained denoiser to reconstruct a residual $\hat{R}_{\theta}$ consistent with the target conditional distribution. The final high-resolution prediction is obtained by adding this sample to the deterministic regression output, i.e., $\hat{Y} = \mu_{\theta} + \hat{R}_{\theta}$.
\begin{figure}[htbp]
    \centering
    \includegraphics[width=0.8\textwidth]{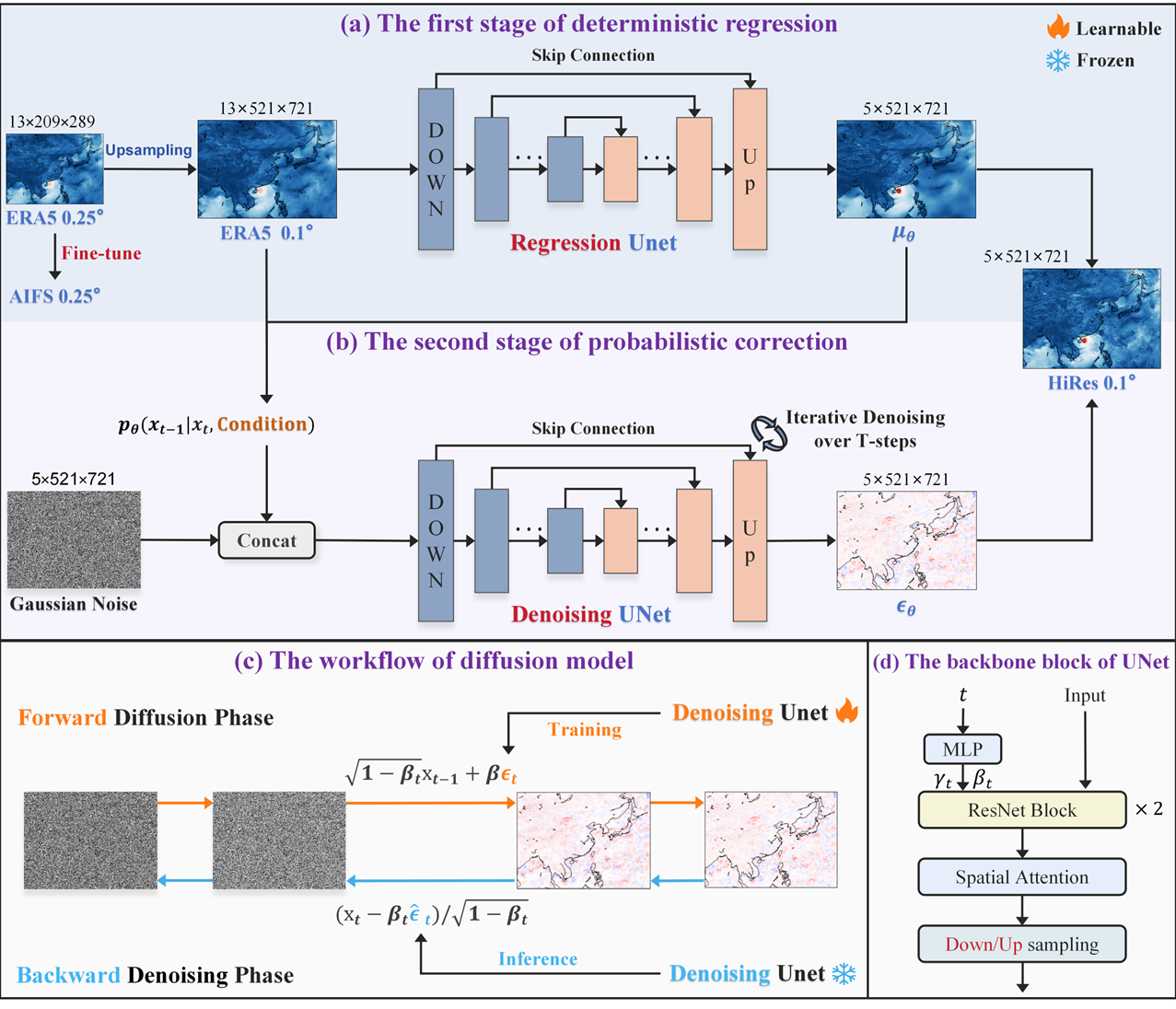}
    \caption{Overall framework of the proposed UNet-Diff. (a) The first stage performs deterministic regression, where the 0.25° ERA5 reanalysis is interpolated and upsampled, followed by a regression UNet for conditional mean estimation. (b) The second stage applies a conditional diffusion model for probabilistic correction, estimating the residual between the true Hires field and the conditional mean field. (c) The workflow of the diffusion process. (d) Backbone of the UNet, composed of ResNet blocks with spatial attention and down/up-sampling layers.}
    \label{fig:fig4}
\end{figure}

\subsection{Training setups}
The UNet-Diff model is trained using the above HiRes reanalysis dataset generated from the Shanghai Typhoon Model over the period from January 2021 to December 2024, with September 2024 withheld as the test set. Among the remaining samples, 80\% are used for training and 20\% for validation. To ensure consistency in spatial resolution, the original 9 km HiRes outputs are interpolated to a regular 0.1° × 0.1° grid. Input features are derived from ERA5 at 0.25°× 0.25° resolution and include both surface-level variables (e.g., 2-m temperature, 10-m wind components, mean sea level pressure, total column water vapor) and mid-tropospheric fields (e.g., geopotential, temperature, wind at 850 and 500 hPa) (see Table 1). Outputs include 2-m temperature, 10-m wind components, mean sea level pressure, and maximum radar reflectivity. As described above, the training process follows a two-stage architecture. In the first stage, a U-Net-based regression model is trained to estimate the deterministic mean field of the high-resolution targets. In the second stage, a conditional diffusion model learns to reconstruct the residuals between the true HiRes fields and the regression output. The diffusion model is conditioned on the concatenated tensor of the regression output and upsampled ERA5 inputs, resulting in a 23-channel input to the denoising network. The model is trained with 1000 sampling steps using a cosine noise schedule and a predv objective in the normalized domain. The full training is conducted on 8 NVIDIA A100 GPUs (40 GB each) using PyTorch’s DistributedDataParallel (DDP) framework. The total training duration is approximately 42 hours for 200 epochs, with mixed precision (AMP) enabled, a learning rate of 2 × 10e-4, AdamW optimizer, batch size of 1 per GPU, and gradient clipping at 0.5. This setup ensures both high forecast skill and operational efficiency, supporting rapid inference with high-resolution structural fidelity.
\section{Results}
\subsection{10-m surface wind}
To assess the performance of different models in reproducing near-surface wind structures, we examine the 10-m wind field at 0000 UTC on 20 September 2024 across the full model domain. As shown in Fig. 5, the ERA5 reanalysis at 0.25° resolution (Fig. 5a) displays a smoother wind field compared to the 0.1°HiRes reanalysis (Fig. 5d), especially over regions with complex terrain, where fine-scale structures are largely absent. For the downscaled results, the UNet model introduces some terrain-induced wind features over land, but overall remains overly smoothed, particularly over the ocean where the absence of orographic forcing limits its ability to reproduce realistic wind structures. In contrast, a single member from the UNet-Diff ensemble demonstrates wind patterns that closely match those of the HiRes reanalysis over both land and ocean, reflecting the diffusion model’s superior capability in recovering high-frequency details. Further evaluation focuses on model performance in typhoon conditions, using Typhoon Yagi (2025) as a case study (Fig. 6). We compare ERA5, UNet trained without typhoon samples (UNetnoTCsamples), UNet, and UNet-Diff against the HiRes reference. The HiRes reanalysis indicates a maximum wind speed of 43.5 m/s, whereas ERA5 significantly underestimates the intensity at 25.5 m/s. Notably, the UNetnoTCsamples model yields a maximum wind speed of only 26.7 m/s, underscoring the importance of including extreme typhoon cases during training—AI models cannot infer physical processes absent from the training distribution. While both UNet and UNet-Diff enhance the typhoon intensity relative to ERA5, UNet-Diff more faithfully reproduces the structural characteristics of the typhoon, indicating its advantage in capturing the intensity and morphology of extreme events.
\begin{figure}[htbp]
    \centering
    \includegraphics[width=0.8\textwidth]{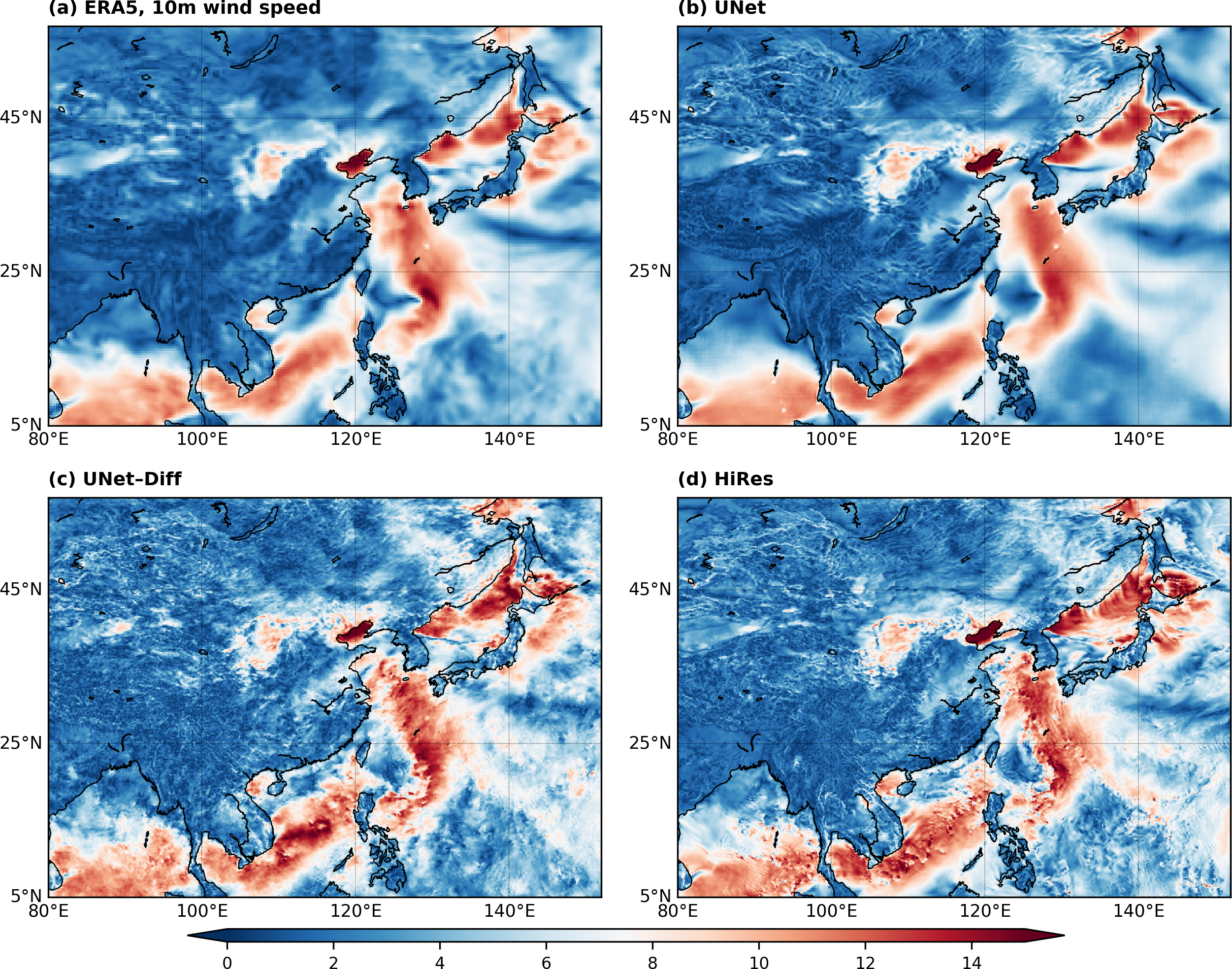}
    \caption{Spatial distributions of 10-m wind speed (unit: m/s, SHTM domain) from (a) ERA5, (b) UNet, (c) UNet-Diff and (d) HiRes (target), at 0000 UTC 20 September 2024.}
    \label{fig:fig5}
\end{figure}

\begin{figure}[htbp]
    \centering
    \includegraphics[width=0.8\textwidth]{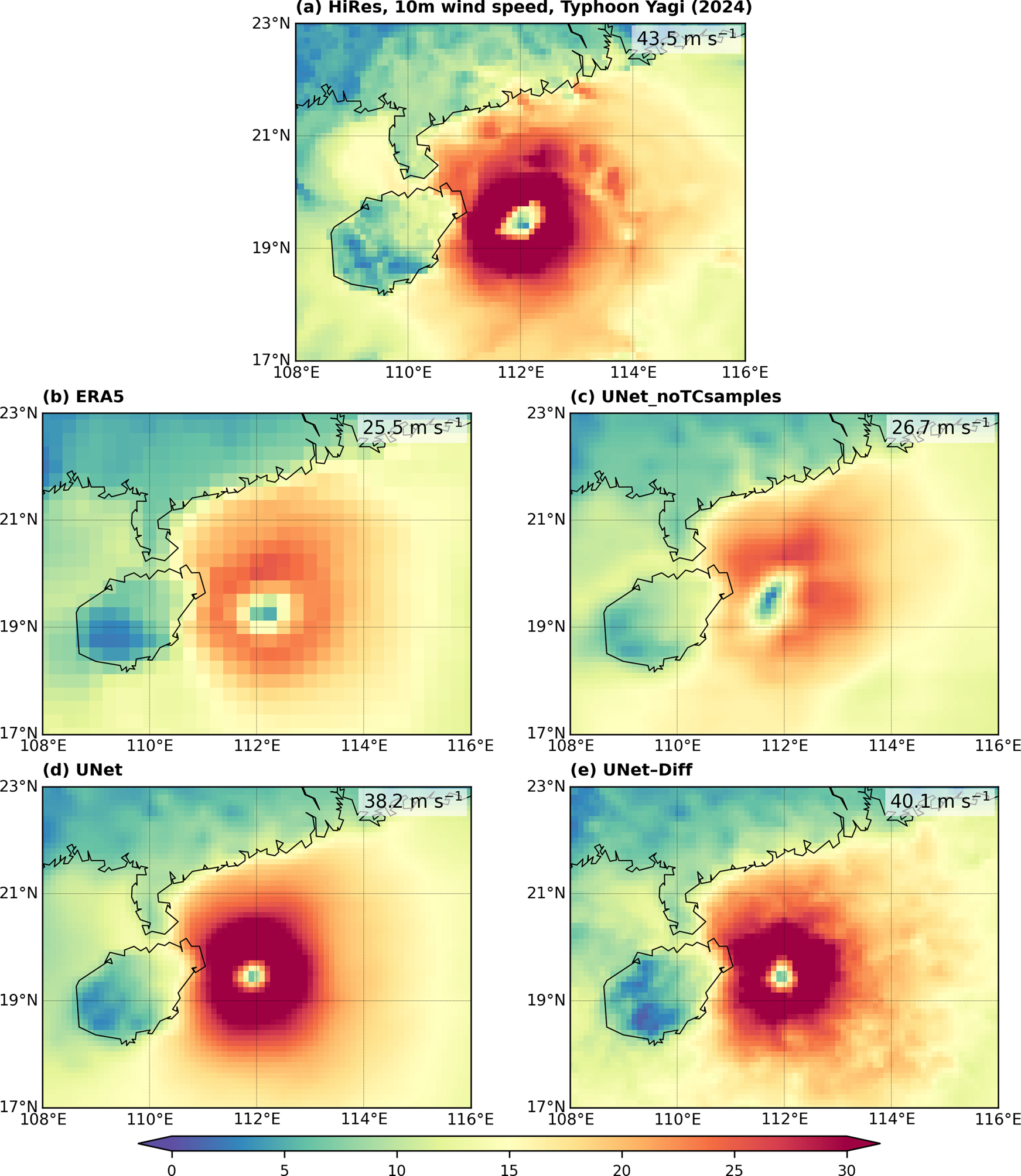}
    \caption{Spatial distributions of 10-m wind speed (unit: m/s, typhoon area) from (a) HiRes, (b) ERA5, (c) UNetnoTCsamples, (d) UNet, (e) UNet-Diff, at 0000 UTC 6 September 2024 during Typhoon Yagi (2024). The maximum 10-m wind speed is annotated in each panel.}
    \label{fig:fig6}
\end{figure}

\subsection{Maximum radar reflectivity}
Since maximum radar reflectivity is not included in the input features, this study additionally trains an observation operator within the downscaling framework to infer convective and typhoon structures. As illustrated in Figure 7, we examine four different time snapshots to compare the downscaled outputs from UNet and UNet-Diff. At most time steps, the deterministic UNet exhibits overly smoothed, fragmented, and poorly organized reflectivity fields, particularly in regions associated with convective clusters. It fails to reproduce fine-scale intense reflectivity cores, indicating the limitations of single-stage regression models in capturing extreme precipitation and organized convection. By contrast, UNet-Diff demonstrates markedly improved spatial organization, reflectivity intensity, and structural detail, producing results that more closely resemble the HiRes reanalysis. This indicates that the diffusion process effectively restores high-frequency details and enforces physical consistency, thereby enhancing both the diversity and fidelity of the generated outputs. Figure 8 evaluates the statistical performance of the two models using HiRes as ground truth. In Figure 8a, probability density functions (PDFs) of maximum reflectivity across the test set show that UNet significantly underestimates the distribution tail beyond 20 dBZ. In contrast, the 20-member UNet-Diff ensemble captures a broader range of reflectivity values, with its mean closely aligning with the HiRes reference, particularly in the 20–45 dBZ range. This demonstrates the model's ability to recover spatial variability and intensity in extreme precipitation events. Figure 8b further assesses the models using threat scores (TS) across multiple reflectivity thresholds. While UNet-Diff slightly underperforms UNet at the >0 dBZ and >10 dBZ thresholds, it substantially outperforms it at thresholds above 20 dBZ. This suggests that UNet-Diff does not optimize toward score maximization via oversmoothing, but instead more accurately captures the distribution of heavy precipitation, offering better guidance for extreme weather and typhoon forecasts.

\begin{figure}[htbp]
    \centering
    \includegraphics[width=0.8\textwidth]{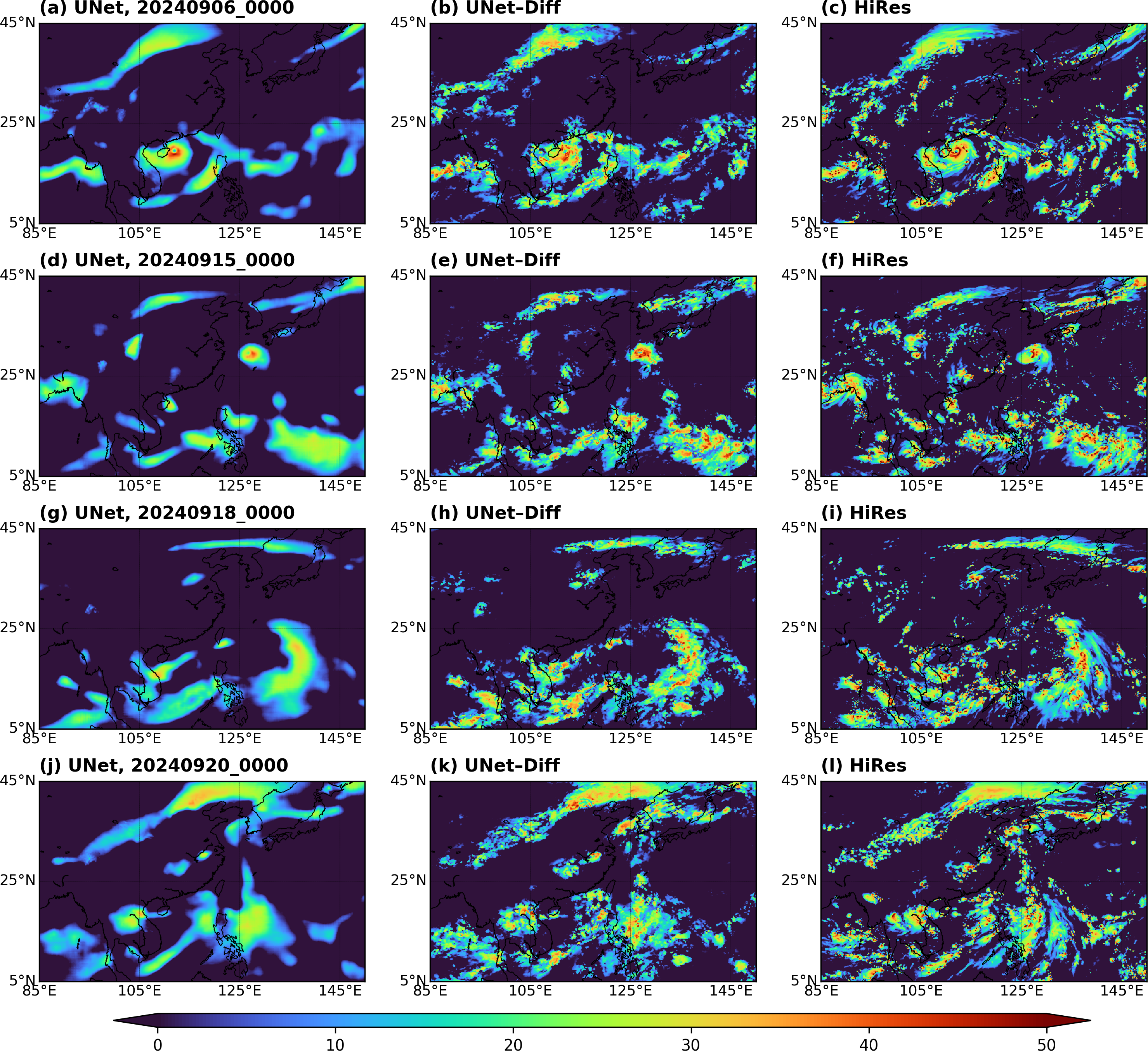}
    \caption{Spatial distributions of maximum radar reflectivity (unit: dBZ) from UNet (left), UNet-Diff (middle) and HiRes (right), evaluated at (a)-(c) 0000 UTC 6 September 2024, (d)-(f) 0000 UTC 15 September 2024, (g)-(i) 0000 UTC 18 September 2024, and (j)-(l) 0000 UTC 20 September 2024, respectively.}
    \label{fig:fig7}
\end{figure}

\begin{figure}[htbp]
    \centering
    \includegraphics[width=0.8\textwidth]{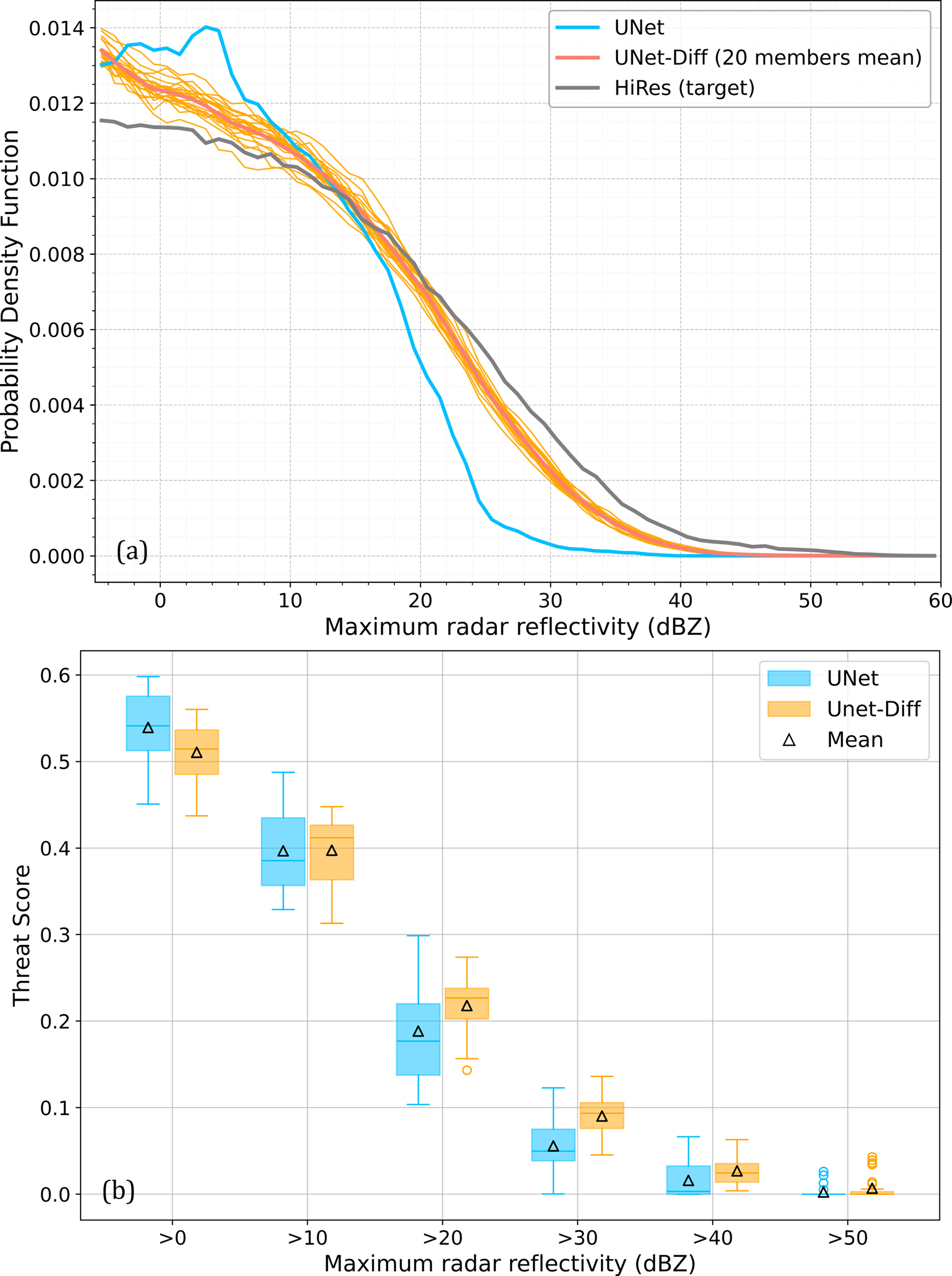}
    \caption{(a) Probability density functions of maximum radar reflectivity (dBZ) for UNet (light blue), UNet-Diff 20 ensemble members (orange thin) and their ensemble mean (salmon), compared with the high-resolution target (gray).(b) Box-and-whisker comparison of Threat Score (TS) for UNet and Unet-Diff forecasts of maximum radar reflectivity at different thresholds.Black triangles mark the mean. PDFs and scores are evaluated from 1 September to 15 September 2024.}
    \label{fig:fig8}
\end{figure}

\subsection{Fune-tune and real-time typhoon forecasts}
In addition to the downscaling experiments driven by ERA5 inputs, we further fine-tune the proposed model using training samples constructed from AIFS forecasts and the corresponding high-resolution outputs of the SHTM hybrid model. The fine-tuning is conducted with data from June 2025, and the model is evaluated using two typhoon cases from July and August. We first assess the near-surface wind field for Typhoon Danas (2025) (Fig. 9). Similar to the ERA5-driven results, the 6-hour forecasts from AIFS show a consistently weak typhoon circulation, while the SHTM outputs exhibit stronger and more realistic wind intensities. Both the UNet and UNet-Diff (one member) models produce enhanced typhoon winds, and notably, the structural characteristics of the UNet-Diff forecast align more closely with SHTM than with the deterministic UNet. This consistency indicates that the proposed two-stage framework maintains robustness and transferability even when switching input sources from reanalysis to AIWP forecasts.
\begin{figure}[htbp]
    \centering
    \includegraphics[width=0.8\textwidth]{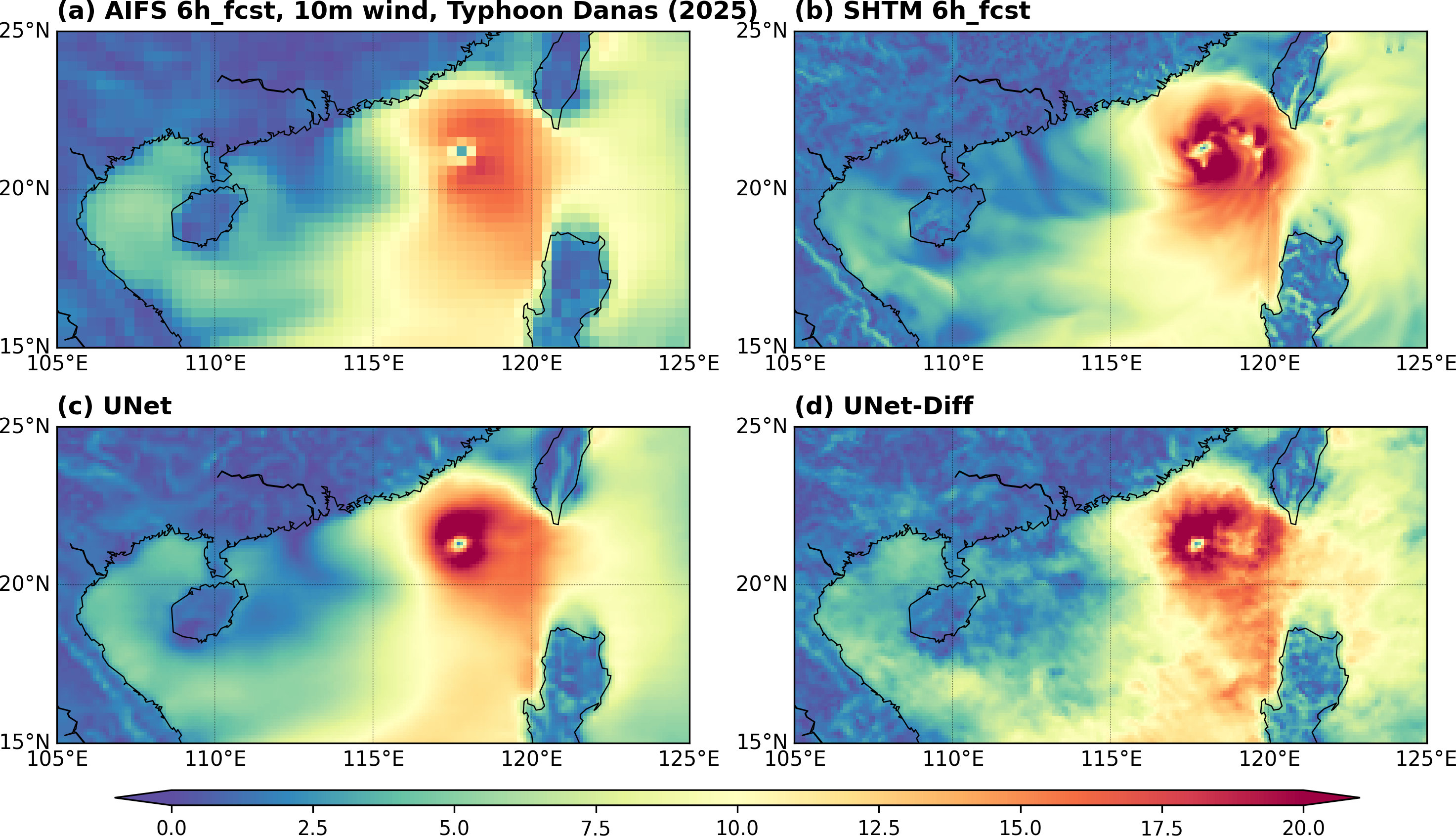}
    \caption{Spatial distributions of 10-m wind (unit: m/s) forecasts for Typhoon Danas (2025) from (a) AIFS 6-h forecasts, (b) SHTM 6-h forecasts, (c) UNet and (d) UNet-Diff predictions, valid at 1200 UTC 5 July 2025.}
    \label{fig:fig9}
\end{figure}

We further examine Typhoon Podul (2025) by evaluating both track and intensity forecasts. As shown, all methods—except for ECMWF-IFS—produce tracks that closely follow the AIFS forecast and are generally consistent with observations. Importantly, the 20 ensemble members of UNet-Diff exhibit a compact spread, and their ensemble-mean track remains close to that of AIFS throughout the forecast horizon. While the UNet-Diff ensemble shows slightly larger track errors during the 0–24 h lead time compared to AIFS, it achieves smaller errors beyond 108 h. These results collectively demonstrate that the proposed downscaling framework, when fine-tuned with AIWP–hybrid model pairs, delivers reliable and transferable performance for real-time typhoon forecasting. The performance in predicting maximum sustained wind speed (Vmax) reveals more pronounced differences (Fig. 10b). AIFS consistently underestimates typhoon intensity throughout the forecast period, which reflects a common weakness in current AIWP models in capturing typhoon intensification. In contrast, both SHTM and UNet-Diff effectively track the increasing intensity, with UNet-Diff showing a more dynamic growth trend that closely mirrors the observed intensification around 84 h. The spread among UNet-Diff members remains moderate, balancing forecast uncertainty and ensemble realism. Overall, Fig. 10 highlights the capability of UNet-Diff to provide skillful typhoon intensity forecasts by effectively emulating the physics-informed SHTM outputs. This demonstrates its utility as a hybrid AI-physics surrogate model capable of improving intensity prediction while maintaining track accuracy.

\begin{figure}[htbp]
    \centering
    \includegraphics[width=0.8\textwidth]{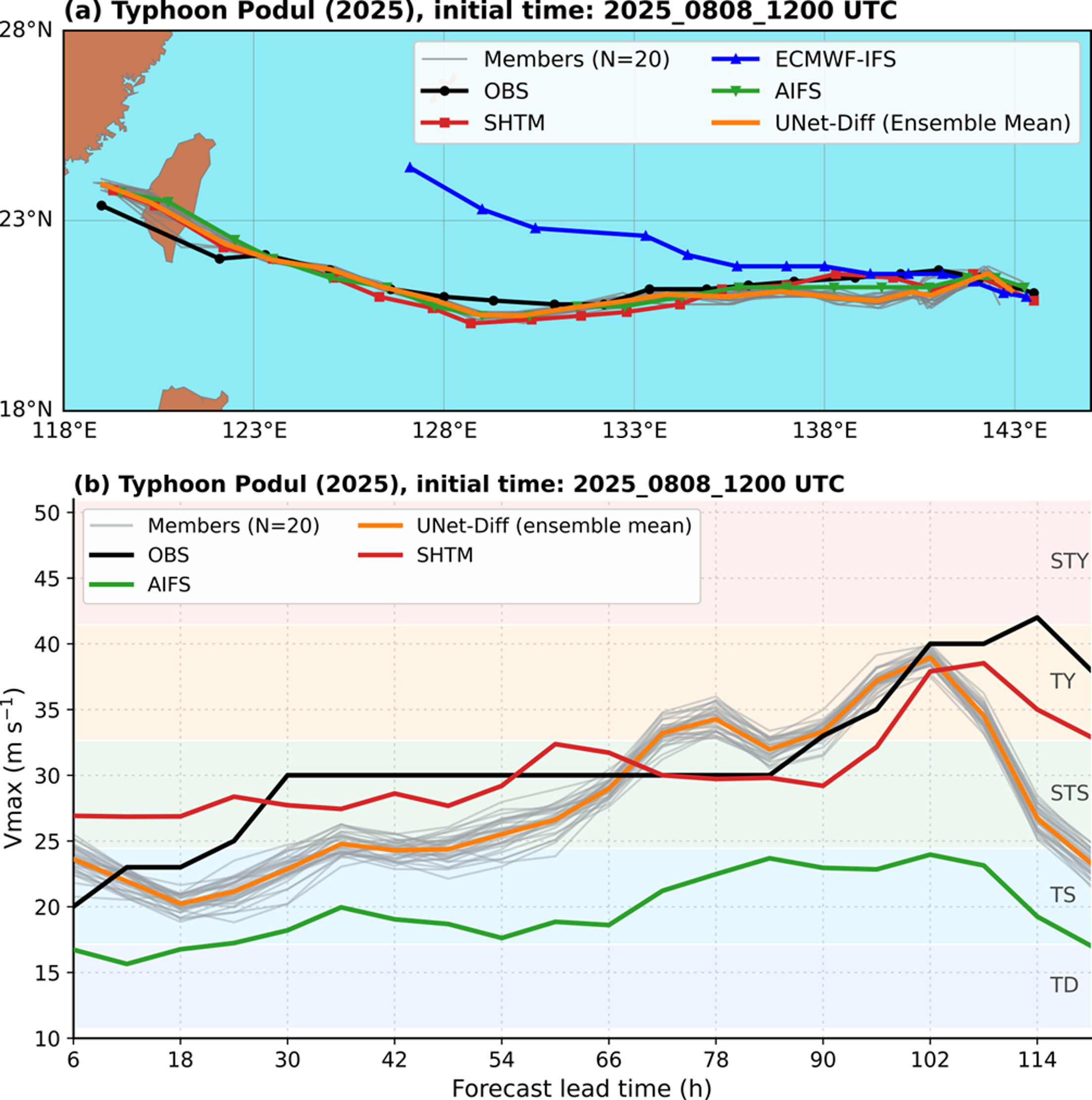}
    \caption{(a) Track and (b) maximum sustained wind speed (Vmax, unit: m/s) forecasts versus lead time (6–120 h), from BABJ observations (black), AIFS (green), SHTM (red), ECMWF-IFS (blue), and Unet-Diff 20 ensemble members (gray) and their ensemble mean (orange), for Typhoon Podul (2025) initialized at 1200 UTC 8 August 2025.}
    \label{fig:fig10}
\end{figure}

\subsection{Computational efficiency}
Table 2 compares the computational efficiency of the proposed UNet-Diff model with the operational SHTM hybrid model. For a 120-hour forecast, the UNet-Diff model requires only 3.0 minutes on a single NVIDIA A100 GPU (sampling timesteps = 50), while the SHTM simulation takes approximately 66.0 minutes on a 2240-core CPU cluster (2.80 GHz). Despite the dramatic reduction in computational cost—over 20× faster inference—UNet-Diff maintains competitive skill in capturing typhoon intensity and structure. This significant speedup, combined with its fidelity in reproducing high-resolution features, highlights the potential of diffusion-based AI surrogates to serve as efficient and accurate alternatives to traditional physics-based models for rapid weather prediction.
\begin{table}[t]
  \centering
  \caption{A list of the input and output details for the UNet--Diff model. 
  Input channels include both single-level and pressure-level variables.}
  \label{tab:io}
  \begin{tabularx}{\linewidth}{%
      l
      >{\raggedright\arraybackslash}X
      >{\raggedright\arraybackslash}X}
    \toprule
    & \textbf{Input} & \textbf{Output} \\
    \midrule
    Horizontal resolution
      & $0.25^{\circ}\!\times\!0.25^{\circ}$
      & $0.1^{\circ}\!\times\!0.1^{\circ}$ \\
    Pixel size
      & $13 \times 209 \times 289$
      & $5 \times 521 \times 721$ \\
    Single-level variables
      & \begin{tabular}[t]{@{}l@{}}
          Mean sea level pressure\\
          2-m temperature\\
          10-m v-component of wind\\
          10-m u-component of wind\\
          Total column water vapor
        \end{tabular}
      & \begin{tabular}[t]{@{}l@{}}
          Mean sea level pressure\\
          2-m temperature\\
          10-m v-component of wind\\
          10-m u-component of wind\\
          Maximum radar reflectivity
        \end{tabular} \\
    Pressure-level variables (850 and 500 hPa)
      & \begin{tabular}[t]{@{}l@{}}
          Geopotential\\
          Temperature\\
          U-component of wind\\
          V-component of wind
        \end{tabular}
      & --- \\ 
    \bottomrule
  \end{tabularx}
\end{table}
\section{Conclusions and Outlook}
This study proposes a two-stage deep learning framework that combines deterministic UNet regression and residual denoising diffusion (UNet-Diff) to downscale coarse-resolution AIWP or reanalysis forecasts into high-resolution typhoon fields. The model is trained using a newly constructed 9-km regional typhoon reanalysis dataset and evaluated on both reanalysis and real-time AIWP forecasts. The key findings are summarized as follows:

1. High-fidelity structural restoration: UNet-Diff effectively reconstructs the fine-scale wind and radar reflectivity structures of typhoons, outperforming baseline UNet models, particularly over the ocean and in regions of intense convection. Its ability to recover high-frequency details stems from the diffusion process, which learns the residual distribution beyond the deterministic mean prediction.

2.	Superior performance in extreme events: The inclusion of typhoon samples in training proves crucial for capturing extreme intensities. UNet-Diff exhibits stronger skill than UNet and ERA5 in predicting typhoon maximum wind speed and spatial structure, aligning more closely with the hybrid physics-based SHTM output.

3.	Transferability and robustness: The model remains robust when fine-tuned on AIWP-SHTM paired data, and generalizes well to real-time forecasts from AIFS. It delivers skillful predictions for typhoon intensity and track, with moderate ensemble spread and consistent alignment with the reference hybrid model.

4.	Computational efficiency: UNet-Diff achieves over 20× faster inference compared to the operational SHTM, requiring only 3 minutes on a single NVIDIA A100 GPU for a 120-h forecast making it well-suited for rapid typhoon prediction in time-sensitive settings.

These results highlight the promise of integrating deep generative models with physical constraints for weather downscaling. UNet-Diff serves as a viable AI surrogate to hybrid numerical models, retaining physical realism while offering significant speedup. Future work will focus on extending this framework to multi-variable and multi-level forecasting. Moreover, incorporating physical priors and consistency constraints—such as mass conservation or balance equations—into the diffusion model may further improve realism and generalization. Real-time operational integration and ensemble-based uncertainty quantification will also be explored to support typhoon risk assessment and early warning systems.

\section{Data and Code Availability}
The pretrained weights for both the UNet and UNet-Diff models used in this study are publicly available. Additionally, the training code for the UNet-Diff model is open-sourced and accessible at \url{https://github.com/ZeyiNiu/SHTM}. The high-resolution reanalysis dataset (HiRes) developed in this study can be obtained upon request from the corresponding author.

\section{Acknowledgments}
 This research was supported by the Special Project-Original Exploration (Grant 42450163); the National Youth Science Foundation of China Project (Grant 4240050560); Research and development of key technologies for artificial intelligence regional typhoon forecasting model.

\nocite{*}

\bibliographystyle{ametsocV6.bst}

\end{document}